\documentclass[pdflatex,sn-nature,twoside]{sn-jnl}
\usepackage{times}
\usepackage{fullpage}
\usepackage{fix-cm}
\usepackage{ulem}
\usepackage{array}
\usepackage{graphicx}
\usepackage{multirow}
\usepackage{amsmath,amssymb,amsfonts}
\usepackage{amsthm}
\usepackage{bbm}
\usepackage[title]{appendix}
\usepackage{xcolor}
\usepackage{textcomp}
\usepackage{manyfoot}
\usepackage{algorithmicx}
\usepackage{algpseudocode}
\usepackage{listings}
\usepackage{xspace}
\usepackage{multicol}
\usepackage{placeins}
\usepackage{booktabs}
\usepackage{siunitx}

\usepackage{mathtools}
\usepackage[capitalise]{cleveref}
\usepackage{caption}
\usepackage{subcaption}
\usepackage{svg}
\usepackage{threeparttable}
\usepackage{acronym}
\usepackage{lineno}
\usepackage{defs}

\graphicspath{{./plots/}}

\DeclareCaptionLabelSeparator{bar}{\space\textbf{\textbar}\space}
\DeclareCaptionFont{captionfont}{\fontsize{11pt}{11pt}\selectfont}
\usepackage{tablefootnote}
\usepackage{threeparttable}

\title{HEPTv2: End-to-End Efficient Point Transformer for Charged Particle Reconstruction}

\author[1]{Siqi Miao}
\equalcont{These authors contributed equally to this work.}

\author[1]{Shitij Govil}
\equalcont{These authors contributed equally to this work.}

\author[2]{Jack P.~Rodgers}
\author[2]{Mia Liu}
\author[3]{Javier Duarte}
\author[4]{Shih-Chieh Hsu}
\author*[4]{Yuan-Tang Chou}\email{yuan-tang.chou@cern.ch}
\author*[1]{Pan Li}\email{panli@gatech.edu}

\affil[1]{School of Electrical and Computer Engineering, Georgia Institute of Technology, Atlanta, GA, USA}
\affil[2]{Department of Physics and Astronomy, Purdue University, West Lafayette, IN, USA}
\affil[3]{Department of Physics, University of California San Diego, La Jolla, CA, USA}
\affil[4]{Department of Physics, University of Washington, Seattle, WA, USA}

\abstract{
Charged-particle tracking---the reconstruction of particle trajectories from sparse detector measurements---is a fundamental inference problem in high-energy physics and a canonical example of learning under extreme combinatorial ambiguity. At the High-Luminosity Large Hadron Collider (HL-LHC), tracking must remain accurate and computationally efficient despite an unprecedented density of simultaneous particle collisions. Existing graph neural network approaches achieve strong performance, but incur substantial computational costs from graph construction and processing, while transformer-based approaches still rely on auxiliary processing stages that prevent end-to-end optimization. To address this, we present HEPTv2, an end-to-end point-transformer architecture that directly reconstructs particle tracks from detector hits in a single trainable pipeline. HEPTv2 combines a locality-aware point encoder with a track decoder that predicts complete trajectories without intermediate graph-building, clustering, or filtering stages. The encoder leverages locality-sensitive hashing in detector coordinate space to preserve tracking-relevant geometric structure while enabling efficient local attention. The decoder resolves local ambiguities through sectorized decoding and direct hit-to-track prediction under joint encoder-decoder supervision, allowing the entire reconstruction pipeline to be optimized end-to-end. On the TrackML benchmark, HEPTv2 achieves a 98.6\% double-majority tracking efficiency at a 0.8\% fake rate, while requiring only $\sim$15~ms inference time and 0.4~GB peak memory per event on a single NVIDIA A100 GPU. Both latency and memory both scale approximately linearly for events containing up to $5\times10^5$ detector hits. HEPTv2 establishes a new state of the art in the accuracy-latency trade-off, improving tracking efficiency by more than 4.5\% over the strongest prior transformer approach and by 1.1--2.2\% over highly optimized graph-based pipelines, while reducing inference latency by factors of 7 and 38--52, respectively.
These results demonstrate for the first time that end-to-end transformer architectures can deliver both the accuracy and computational efficiency required for real-time particle reconstruction at the HL-LHC.
}

\begin{document}
\maketitle

\acresetall

\begin{figure}[t]
\vspace{-2mm}
    \centering
\includegraphics[trim={1.2cm 0 0.0cm 0 0},clip,
        width=0.99\linewidth,
    ]{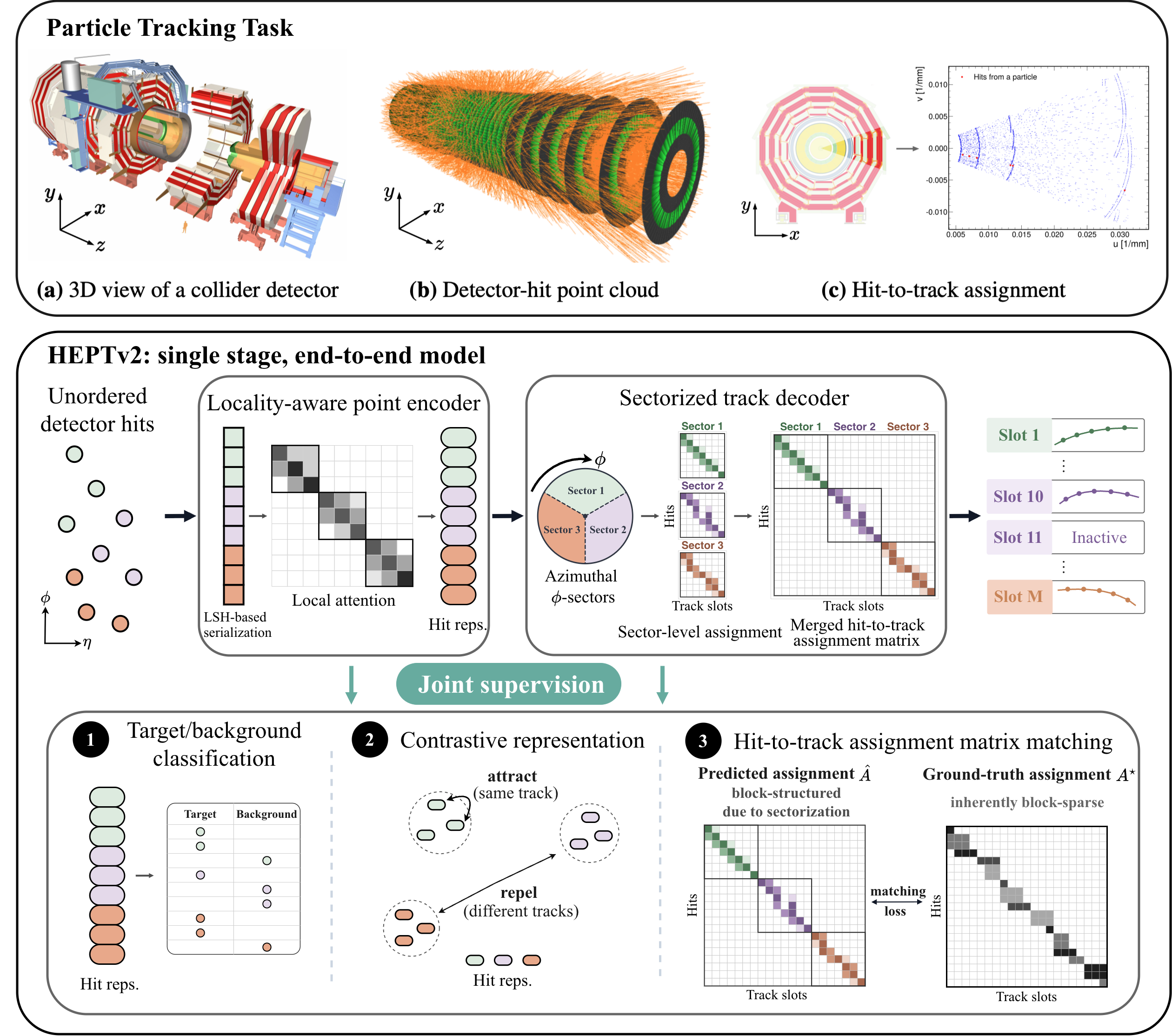}
    \vspace{1mm}
\caption{
Overview of charged-particle tracking and the HEPTv2 pipeline. Top, adapted in part from~\cite{ju2021performance,miao2024locality}, a collider detector records sparse measurements, a high-pileup event produces a large unordered hit cloud, and reconstruction associates hits into particle trajectories. Bottom, our HEPTv2 pipeline uses a locality-aware point encoder and a sectorized track decoder with joint end-to-end supervision.
}
    \label{fig:heptv2}
    \vspace{-6mm}
\end{figure}


\section{Introduction}

Data-intensive scientific instruments increasingly require algorithms that can reconstruct latent physical objects from large, noisy, and unordered sets of measurements~\cite{kiehn2019trackml, pata2021mlpf, shlomi2021graph}.
In \ac{HEP}, a central example is charged-particle tracking at the \ac{LHC}, where silicon detectors record sparse spatial hits as charged particles traverse successive detector layers, and reconstruction algorithms must associate the hits produced by each particle into a trajectory~\cite{duarte2022graph}, as illustrated in Figure~\ref{fig:heptv2}. 
Accurate reconstruction of these trajectories is a backbone of collider event reconstruction, providing essential inputs to precision measurements and searches for physics beyond the \ac{SM}~\cite{strandlie2010track,dezoort2021charged}. 
The scale of this task will increase sharply at the \ac{HL-LHC}, where the average number of simultaneous proton--proton interactions per bunch crossing (pileup) is expected to rise from a typical Run-3 leveling target of 64 to as many as 200, increasing detector occupancy and the combinatorial ambiguity of hit association~\cite{aberle2020high,kiehn2019trackml,atlas2019expected}. 
Tracking algorithms must therefore preserve high reconstruction efficiency and low fake rates while satisfying stringent throughput and latency requirements, making charged-particle tracking a central algorithmic and computing challenge for next-generation collider experiments~\cite{atlas2024software,lantz2020speeding,elabd2022graph, gessinger2025traccc}.

The difficulty lies in the combinatorial structure of the problem: although the physically meaningful hit associations are sparse and local, as constrained by detector geometry and magnetic-field-induced curvature, the reconstruction algorithm must still identify them within a large, noisy set of detector hits~\cite{strandlie2010track,amrouche2019tracking}.
Traditionally, this problem has been addressed with classical pattern-recognition pipelines, most prominently \ac{CKF}, which exploit detector geometry and magnetic-field-constrained particle motion to build track candidates from local measurements~\cite{fruhwirth1987application,strandlie2010track}. 
Although highly effective, these pipelines remain difficult to accelerate efficiently on modern parallel hardware---their track building process involves heavy candidate branching, dynamic bookkeeping, and stage-wise hit propagation, all of which are challenging to map onto highly parallel execution.
Under \ac{HL-LHC} pile-up conditions, this makes classical track building a major computing bottleneck~\cite{farrell2017hep,bapst2020pattern,abidi2022charged,atlas2024software} and has driven substantial efforts to redesign or accelerate the \ac{CKF} workflow~\cite{lantz2020speeding,atlas2024software}.

This bottleneck has also motivated learning-based alternatives, particularly \acp{GNN},
which represent detector hits as nodes and candidate local associations as edges, thereby directly exploiting the sparse relational structure of the tracking problem~\cite{dezoort2021charged,ju2021performance,elabd2022graph}. 
The state-of-the-art \ac{GNN}-based pipelines can achieve tracking quality comparable to standard \ac{CKF}-based reconstruction on realistic \ac{HL-LHC}-like benchmarks, while being more amenable to parallel execution on heterogeneous hardware~\cite{ju2021performance,atlas2024computational,lieret2023high}. 
However, in practical deployments, the end-to-end runtime of these methods is often dominated not by GNN inference itself, but by auxiliary pipeline stages such as graph construction, edge pruning, and graph segmentation. These stages introduce data-dependent sparse computation and substantial graph-processing overhead, which become the primary performance bottleneck and complicate efficient deployment on modern accelerators~\cite{lazar2023accelerating,elabd2022graph,atlas2024computational}.

By contrast, transformer models offer a complementary route because their attention layers are highly parallelizable, supported by optimized hardware kernels, and do not require explicit graph-based processing. 
Nonetheless, naively applying global self-attention to tracking is impractical: it scales quadratically with the number of hits and would devote most of its computation to unrelated hit pairs~\cite{vaswani2017attention}. 
To reduce this cost, efficient attention methods developed for language, images, and 3D scenes introduce structured sparsity and locality~\cite{zaheer2020big,liu2021swin,liu2023flatformer}---for example, serializing the input point set into a one-dimensional sequence with hand-crafted ordering rules and applying local attention over it~\cite{wu2024point}. 
However, tracking detectors have irregular geometry and non-uniform hit density, so predefined locality patterns can either separate physically relevant neighbors or waste computation on unrelated hits that happen to be adjacent in the imposed ordering. HEPT~\cite{miao2024locality} addressed this with \ac{LSH}, serializing the unordered point set into one-dimensional sequences in which spatially nearby points tend to stay close. This enables the use of local blockwise attention with linear complexity while retaining the needed geometric inductive bias for tracking.
However, \ac{HEPT} does LSH in its high-dimensional latent space and still requires an external clustering stage such as \ac{DBSCAN}~\cite{ester1996density} to convert learned per-hit embeddings into final tracks. It was also evaluated through the separability of its learned hit embeddings rather than as an end-to-end tracking method under standard physics metrics like tracking efficiency and fake rate~\cite{miao2024locality}. This leaves a gap toward a deployable pipeline: The clustering stage is not optimized jointly with the model, may introduce fake or fragmented tracks, and can limit  throughput~\cite{lieret2023high}.

To close this gap, we introduce HEPTv2, which, to the best of our knowledge, is the first single-stage, end-to-end efficient point transformer for charged-particle tracking. 
HEPTv2 couples a locality-aware point encoder with a direct track-construction decoder, so that the final tracks can be efficiently predicted within a single trainable pipeline, eliminating the need for external clustering or graph processing.
On the encoder side, HEPTv2 builds on \ac{HEPT} by adopting LSH-based serialization for efficient encoding~\cite{miao2024locality}, but applies it directly in detector coordinate space rather than in the latent space used by \ac{HEPT}, to more effectively capture the geometric locality structure of particle trajectories.
On the decoder side, HEPTv2 directly predicts hit-to-track assignments. 
This is more challenging than decoding in many computer-vision tasks because local neighborhoods in tracking can contain greater ambiguity: many spatially nearby detector hits may belong to different trajectories~\cite{carion2020end,cheng2022masked,schult2023mask3d}. HEPTv2 addresses this challenge in two complementary ways. First, a sectorized decoder resolves assignments within smaller, spatially coherent regions, reducing the ambiguity of each local decoding problem and making direct decoding tractable at the full-event scale. Second, the encoder and decoder are trained jointly under a unified objective, so the encoder learns trajectory-informed, discriminative per-hit representations and the decoder can better separate nearby hits from different trajectories.

We evaluate HEPTv2 on TrackML~\cite{amrouche2019tracking,kiehn2019trackml}, the community-standard benchmark for charged-particle tracking. Released as a public Kaggle challenge, it models a generic HL-LHC-era silicon tracker and underlies essentially all of the baselines we compare against. On this benchmark, HEPTv2 achieves 98.6\% \ac{DM} tracking efficiency, a fake rate of 0.8\%, and an average inference latency of about 15~ms per event on a single NVIDIA A100 GPU. Relative to the strongest prior multi-stage transformer baselines~\cite{miao2024locality,van2025transformers}, it improves \ac{DM} by more than 4.5\% while reducing latency by over $7\times$. It also markedly improves the \ac{DM}--latency trade-off over graph-based pipelines: versus OC-GNN~\cite{lieret2023high}, $+2.2\%$ \ac{DM} at $38\times$ lower latency; versus the highly optimized ACORN-\ac{GNN} pipeline~\cite{ju2021performance,atlas2024computational}, $+1.1\%$ \ac{DM} at $52\times$ lower latency. These results show that HEPTv2 delivers strong tracking quality and low latency simultaneously, a practical step toward deployable tracking under HL-LHC-like constraints. More broadly, they suggest that end-to-end, locality-aware transformers may offer a useful foundation for future learning-based reconstruction in \ac{HEP}.

\begin{figure}[t]
    \centering

    \begin{minipage}[t]{0.66\textwidth}
        \vspace{0pt}
        \centering
        \scriptsize
        \setlength{\tabcolsep}{3pt}
        \resizebox{\linewidth}{!}{%
        \begin{tabular}{lcccc}
            \toprule
            Model 
            & $\epsilon^{\mathrm{DM}}_{\pt>0.9}$  
            & $f_{\pt>0.9}$ 
            & Latency (ms) 
            & Memory (GB) \\
            \midrule
            OC-GNN    & 96.4\% & 0.9\% & 571.5 & 5.4 \\
            ACORN-GNN & 97.5\% & 0.9\% & 783.7  & 16.6 \\
            \midrule

            HEPT+DBSCAN           & 89.6\% & 3.3\% & 105.5 & 7.6 \\
            two-stage MF$^\ddagger$ & 94.1\% & 0.7\% & 99   & --  \\
            \midrule
            HEPTv2 & 98.6\% & 0.8\% & 15.1 & 0.4 \\
            \bottomrule
        \end{tabular}
        }
    \end{minipage}
    \hfill
    \begin{minipage}[t]{0.335\textwidth}
        \vspace{-0.4cm} 
        \centering
        \includegraphics[width=\linewidth]{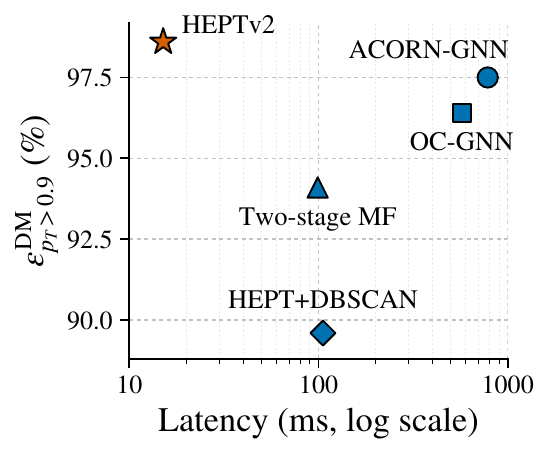}
    \end{minipage}

    \caption{Left: Comparison of HEPTv2 with representative baselines on TrackML. Right: DM--latency trade-off for all compared methods. $^\ddagger$Results are taken from~\cite{van2025transformers} and correspond to the slightly easier setting $\pt>1.0$~GeV, $|\eta|<4$; all other methods use $\pt>0.9$~GeV, $|\eta|<4$.}
    \label{fig:main-results}
    \vspace{-3mm}
\end{figure}

\section{Results}
\label{sec:results}

\subsection{Overview of HEPTv2}\label{sec:overview}

Figure~\ref{fig:heptv2} illustrates the overall architecture of HEPTv2. The model takes an unordered set of detector hits as input and reconstructs charged-particle tracks in a single end-to-end pipeline.

The first module is a locality-aware point transformer encoder that processes the full event. 
It applies LSH-based serialization in $(\eta, \phi)$ space to sort detector hits into one-dimensional sequences.
Because the locality-preserving property of LSH makes hits that are nearby in \((\eta,\phi)\) likely to remain close in the serialized order, efficient local-window attention is then applied over the resulting 1D sequences while preserving the geometric neighborhoods relevant for tracking.
The encoder, therefore, aggregates local geometric context across the full event and outputs contextualized per-hit representations.

\begin{figure}[t]
  \centering
  \includegraphics[width=0.95\textwidth]{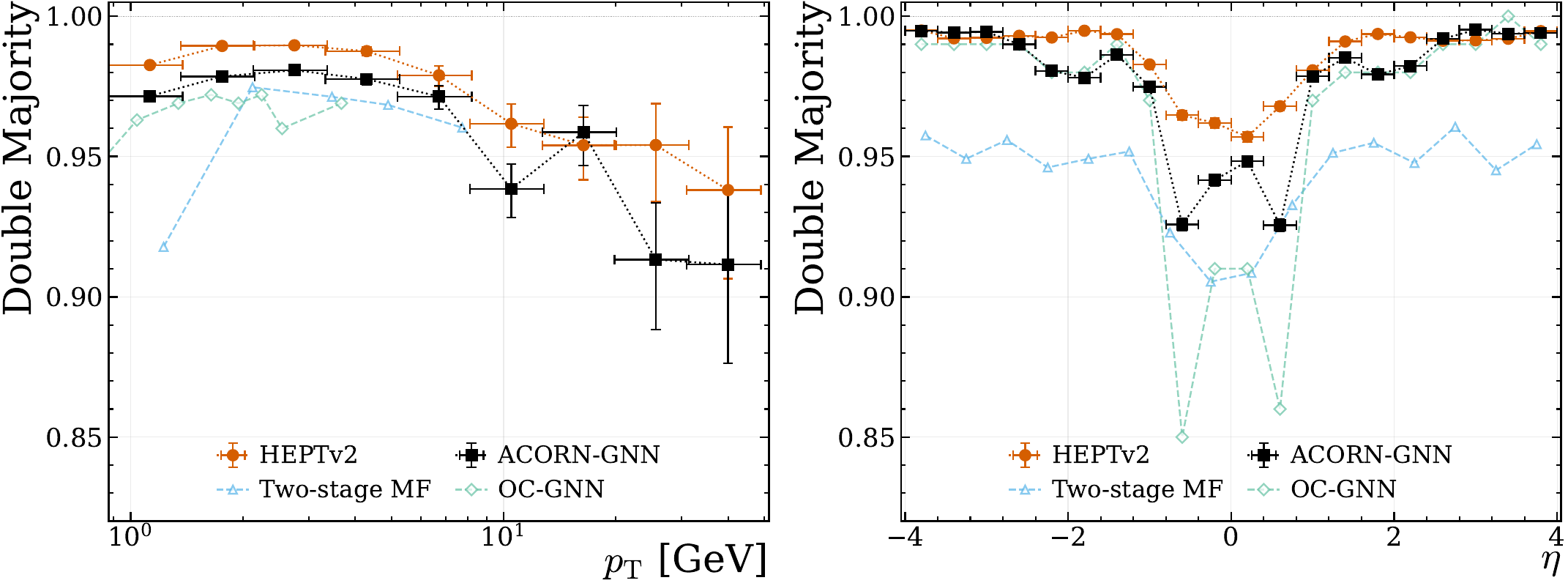}\\[3mm]
  \includegraphics[width=0.95\textwidth]{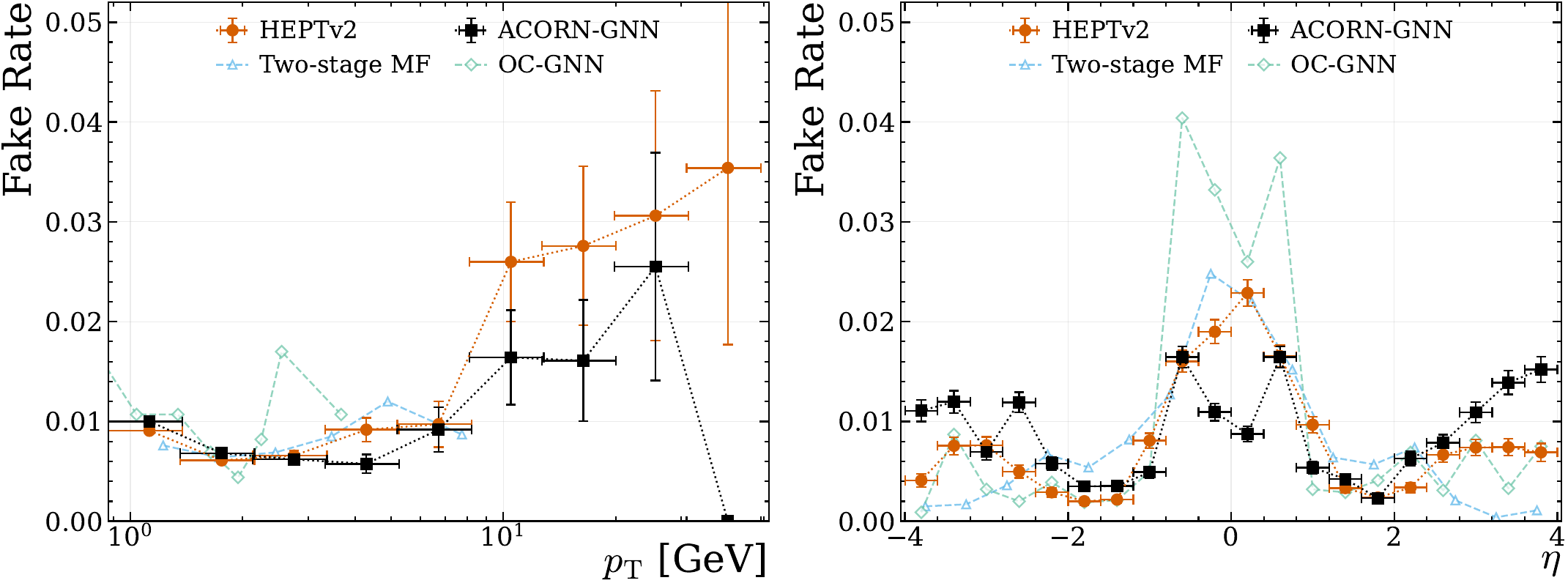}
\caption{Tracking quality across kinematic regimes on TrackML for tracks with \(\pt > 0.9~\mathrm{GeV}\) and \(|\eta| < 4\). DM (top) and fake rate (bottom) are plotted versus \(\pt\) (left) and \(\eta\) (right) for HEPTv2 and baselines. The two-stage MF and OC-GNN values are transcribed from their respective papers.}
  \label{fig:dm-fake-rate}
  \vspace{-3mm}
\end{figure}

The second module is a direct track-construction transformer decoder. 
Conceptually, the decoder maintains a fixed bank of \(M\) learnable latent vectors in \(\mathbb{R}^{d}\), each representing one track slot, where \(M\) is chosen to exceed the maximum expected number of reconstructable tracks in the events of interest. 
For an event with \(N\) detector hits, the decoder predicts an \(M \times N\) hit-assignment matrix, in which each row corresponds to one possible track slot, each column corresponds to one detector hit, and each entry gives a learned assignment score for associating a hit with a track slot.
Hits assigned to the same track slot therefore form a reconstructed track.
These $M$ learnable track slots are not tied to any specific event, but are learned in a data-driven manner to capture generic trajectory patterns rather than fixed track identities. 
They generalize across events through cross-attention to the encoded hit embeddings, so that each slot specializes to the subset of hits that best supports one candidate trajectory and produces the corresponding hit-assignment scores. 
In any given event, only a subset of slots becomes active, while the remainder are learned to stay inactive, allowing the number of active slots to flexibly adapt to the number of tracks present.

Directly predicting the full-event \(M\times N\) hit-to-track assignment is generally more challenging in charged-particle tracking than mask decoding in many computer-vision tasks. 
First, realistic tracking events may contain thousands of reconstructable trajectories, whereas vision decoders typically operate with only a few hundred object slots. Second, local neighborhoods in tracking contain greater ambiguity and noise, because hits that are nearby in detector space may belong to different trajectories rather than to a single object instance, as is more typical in vision settings.
HEPTv2 addresses these challenges in two complementary ways. 
First, it exploits the approximate block-sparse structure of the global hit-to-track assignment, as illustrated in Figure~\ref{fig:heptv2} (bottom right): 
because charged-particle trajectories are approximately local in azimuth, the ground-truth hit assignments for a given track are typically concentrated within a certain \(\phi\) region. Therefore, rather than solving one full \(M\times N\) assignment problem over the entire event, HEPTv2 partitions the event into \(k\) broad \(\phi\)-sectors and solves \(k\) smaller assignment problems of size \(M\times (N/k)\), before merging the sector-level predictions into the final event-level reconstruction. 
Second, HEPTv2 is trained end-to-end so that the encoder and decoder are optimized for the same final reconstruction task. 
Auxiliary supervision on the encoder encourages the model to distinguish target hits from background noise through a classification loss and to produce hit representations that better separate hits from different particle trajectories through an InfoNCE loss~\cite{oord2018representation}.
At the same time, the decoder is trained directly to predict the final hit-assignment matrix under supervision from the ground-truth track assignments via matching-based losses~\cite{schult2023mask3d}.
The full objective is optimized as a weighted sum of these loss terms, so that the encoder learns more discriminative hit representations and cleaner target-background separation, which in turn provide the decoder with more informative evidence for track construction. Further architectural, training, and implementation details are provided in Section~\ref{sec:methods:heptv2}.



\subsection{Experimental Setup and Evaluation Protocol}


We evaluate HEPTv2 on the TrackML dataset~\cite{amrouche2019tracking,kiehn2019trackml}, a widely used high-energy physics benchmark for charged-particle track reconstruction. TrackML simulates a generic silicon tracking detector in an \ac{HL-LHC}-like environment, with a detector model inspired by the ATLAS and CMS tracker upgrades. It provides a useful testbed for learning-based methods motivated by future collider reconstruction challenges. Following recent learning-based tracking studies~\cite{lieret2023high,miao2024locality,van2025transformers}, we use the pixel-detector subset in our experiments, enabling direct comparison with prior learning-based approaches. This setting retains the core hit-association and combinatorial challenges of charged-particle tracking.

We compare HEPTv2 with four representative baselines spanning both graph-based and transformer-based approaches for tracking.
Among graph-based methods, ACORN-GNN~\cite{ju2021performance,atlas2024computational} represents a highly optimized edge-classification pipeline and underlies the ATLAS GNN4ITk reconstruction chain, whose tracking performance has been shown to be comparable to the standard CKF-based workflow in HL-LHC-like settings; it therefore serves as a strong practical graph-based baseline.
OC-GNN~\cite{lieret2023high} provides a complementary graph-based comparison through the object-condensation paradigm, in which graph-based message passing is used to learn more separable per-hit representations and track candidates are subsequently obtained by clustering those representations.
Among transformer-based methods, two-stage MF~\cite{van2025transformers} is included as a strong multi-stage baseline with competitive TrackML performance and explicit latency measurements, whereas HEPT+DBSCAN~\cite{miao2024locality} is the immediate precursor to HEPTv2 and provides a comparison for demonstrating the effect of replacing post-hoc clustering with direct end-to-end track construction.

\subsection{Tracking Performance on TrackML}

Figure~\ref{fig:main-results} shows that HEPTv2 achieves the strongest overall reconstruction quality on TrackML, reaching a \ac{DM} efficiency of 98.6\% with a fake rate of 0.8\%. 
Relative to ACORN-GNN, the strongest graph-based baseline, HEPTv2 improves \ac{DM} by 1.1\% while also reducing fake rate from 0.9\% to 0.8\%. 
The margin is larger against OC-GNN, with a absolute gain of 2.2\% in DM efficiency. 
The gains are even more significant when comparing with transformer-based baselines.
Relative to two-stage MF, HEPTv2 improves \ac{DM} by 4.5\% under a even more challenging setting with selection cut \(\pt > 0.9\)~GeV and \(|\eta| < 4\). This highlights the benefits of LSH-based serialization together with end-to-end joint optimization, in contrast to the simple geometry-based point ordering  with decoupled training stages.
The comparison with HEPT+DBSCAN further demonstrates the benefit of replacing post-hoc clustering with direct end-to-end track decoding, where HEPTv2 raises DM efficiency from 89.6\% to 98.6\% and reduces the fake rate from 3.3\% to 0.8\%.

The kinematic breakdown in Fig.~\ref{fig:dm-fake-rate} shows that the improvement of HEPTv2 is not confined to a narrow region of phase space. As a function of \(\pt\), HEPTv2 maintains the highest \ac{DM} efficiency over most of the measured range and remains more stable than the baselines in the highest-\(\pt\) bins, where the degradation of ACORN-GNN and two-stage MF becomes more pronounced. As a function of \(\eta\), HEPTv2 also achieves the strongest overall \ac{DM} profile across the detector acceptance. Although all methods exhibit a dip near central \(\eta\) due to the ambiguity of hit configurations in this region, HEPTv2 degrades substantially less than the other methods and remains consistently above ACORN-GNN around \(\eta \approx 0\). 
The corresponding fake-rate curves indicate that the remaining errors of HEPTv2 are concentrated mainly in the most challenging kinematic regimes. In particular, the fake rate increases at high \(\pt\) and near central \(\eta\), where local ambiguity is greatest. 
Even so, HEPTv2 remains more robust than two-stage MF and OC-GNN in these regions. Compared with ACORN-GNN, HEPTv2 exhibits a slightly higher fake rate in some regimes, but this trade-off is accompanied by clear gains in \ac{DM} efficiency.
Taken together, these results demonstrate that HEPTv2 delivers the strongest overall reconstruction quality while maintaining robustness across difficult kinematic regimes.

\begin{figure}[t]
\vspace{-4mm}
  \centering

  \includegraphics[width=0.495\linewidth]{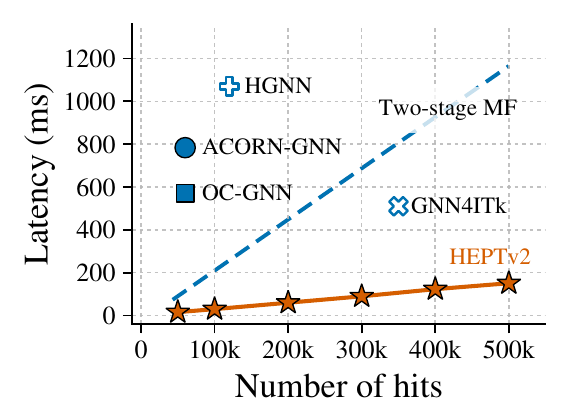}
  \hfill
  \includegraphics[width=0.495\linewidth]{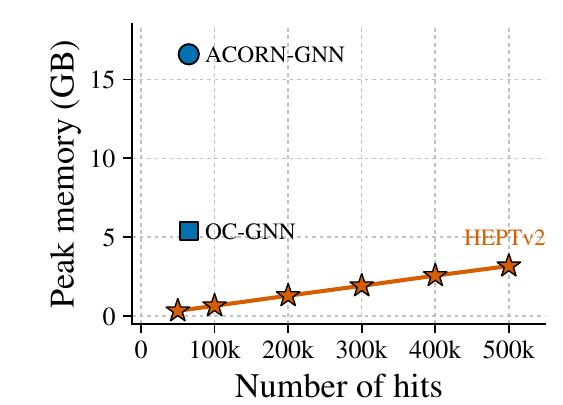}
  \vspace{1mm}
  \caption{
 Scalability comparison between HEPTv2 and baselines.
Left: end-to-end inference latency versus the number of hits.
Right: peak allocated GPU memory versus the number of hits.
Results for HGNN~\cite{liu2026hierarchical}, GNN4ITk~\cite{atlas2024computational}, and two-stage MF~\cite{van2025transformers} are taken from the respective papers; all methods shown were evaluated on a single NVIDIA A100 GPU, but note that HGNN and GNN4ITk were reported under different tracking settings with different event sizes.
  }
  \label{fig:scale}
  \vspace{-3mm}
\end{figure}

\subsection{Latency and Memory Usage}
Beyond tracking quality, HEPTv2 reaches an absolute computational operating point relevant to deployment. It reconstructs a full TrackML event in 15.1 ms using 0.4 GB of peak GPU memory, and this footprint scales near-linearly with event size — to about 150 ms and just above 3 GB at $5\times10^5$ hits (Fig.~\ref{fig:scale}). The sub-GB memory in particular means that full-event reconstruction fits comfortably within the budget of modest or shared accelerators, a practical prerequisite for online use that is not captured by relative speed-ups alone. Against this backdrop, HEPTv2 also dominates the DM–latency trade-off: latency is reduced by $6.6\times$ relative to two-stage MF, and by $7.0\times$ and $51.9\times$ relative to HEPT+DBSCAN and ACORN-GNN, respectively.

For the scalability comparison in Fig.~\ref{fig:scale}, the baseline points for HGNN~\cite{liu2026hierarchical}, GNN4ITk~\cite{atlas2024computational}, and two-stage MF~\cite{van2025transformers} are taken from their original papers, whereas the ACORN-GNN and OC-GNN points are measured by us. All methods were evaluated on a single NVIDIA A100 GPU, but note that HGNN and GNN4ITk were reported under different tracking settings and event sizes.




\subsection{Ablation Studies} 

\begin{figure}[t]
  \centering
  \begin{subfigure}[t]{0.45\linewidth}
    \centering
    \includegraphics[width=\linewidth]{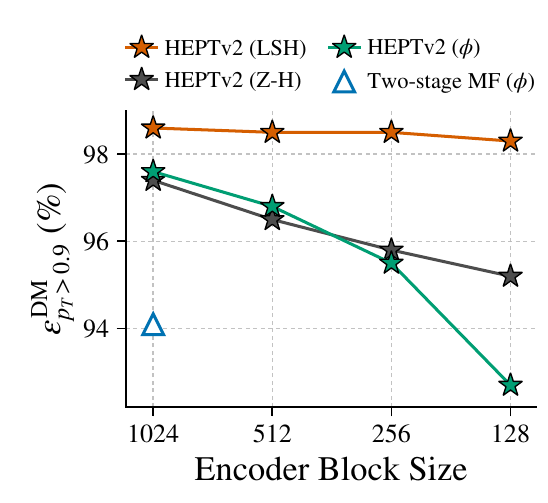}
  \end{subfigure}
  \hfill
  \begin{subfigure}[t]{0.45\linewidth}
    \centering
    \includegraphics[width=\linewidth]{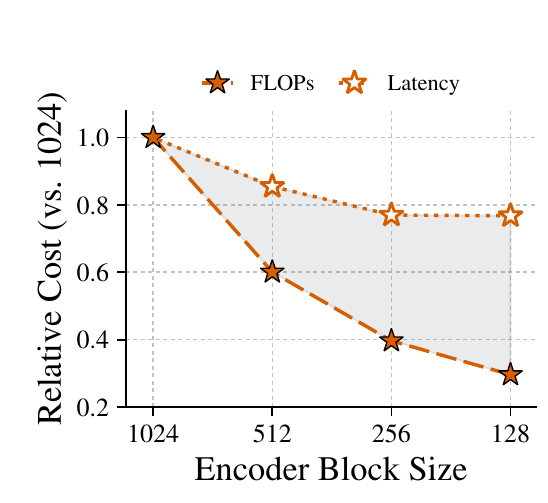}
  \end{subfigure}
\caption{
Effect of the encoder block size on accuracy and encoder cost.
Left: DM for HEPTv2 variants and two-stage MF across encoder block sizes.
Right: encoder FLOPs and measured encoder latency for HEPTv2, each normalized by the corresponding value at block size 1024.
}
\vspace{-1mm}
\label{fig:block_size}
\end{figure}

\begin{table}[t]
\vspace{-6mm}
  \centering
  \caption{Ablation of decoder sectorization and encoder supervision.}
  \label{tab:ablation}
  \begin{tabular}{lccccc}
    \toprule
    Setting & \# $\phi$ sectors & Overlap & Encoder sup. & DM & Fake rate \\
    \midrule
    \multicolumn{5}{l}{\textit{Number of decoder sectors}} \\
    Full-event decoder & 1 & -- & \checkmark & 96.9\% & 0.7\% \\
    2-sector decoder & 2 & -- & \checkmark & 98.2\% & 0.8\% \\
    3-sector decoder & 3 & -- & \checkmark & 98.6\% & 0.8\% \\
    4-sector decoder & 4 & -- & \checkmark & 98.5\% & 0.9\% \\
    5-sector decoder & 5 & -- & \checkmark & 98.5\% & 0.9\% \\
    \midrule
    \multicolumn{5}{l}{\textit{Additional ablations}} \\
    3-sector decoder w/o encoder sup. & 3 & -- & -- & 96.3\% & 0.7\% \\
    3-sector decoder w/ overlap & 3 & \checkmark & \checkmark & 98.6\% & 0.8\% \\
    \midrule
    HEPTv2 & 3 & -- & \checkmark & 98.6\% & 0.8\% \\
    \bottomrule
  \end{tabular}
  \vspace{-2mm}
\end{table}



\vspace{1mm}
\noindent\textbf{Encoder serialization and block size.}
Figure~\ref{fig:block_size} examines how encoder block size affects reconstruction quality, while also comparing the serialization strategy used to define local attention windows. In the left panel, HEPTv2 (LSH) denotes the default model with LSH-based serialization; HEPTv2 (\(\phi\)) replaces LSH with simple azimuthal ordering; and HEPTv2 (Z-H) replaces LSH with the fixed Z-order/Hilbert-curve serialization used in Point Transformer V3~\cite{wu2024point}. 
As the encoder block size is reduced from the default value of 1024 to 128, HEPTv2 (LSH) remains remarkably stable, with only a minor drop in \ac{DM}. By contrast, the fixed-ordering variants degrade much more substantially. 
Specifically, the \(\phi\)-ordered variant performs reasonably at large block sizes, indicating that coarse azimuthal ordering can preserve useful locality when the attention window is wide, but its accuracy deteriorates sharply as the window shrinks. The Z-H variant is more robust than pure \(\phi\)-ordering, suggesting that finer geometric orderings preserve locality better, yet it still degrades noticeably relative to HEPTv2 (LSH).
This comparison indicates that LSH-based serialization preserves tracking-relevant locality more faithfully than fixed orderings, enabling substantially smaller attention blocks without a comparable loss in reconstruction quality.
The right panel of Figure~\ref{fig:block_size} further isolates the encoder of HEPTv2 and sweeps only its block size, with both FLOPs and latency normalized to the corresponding values at the default block size of 1024. Under this setting, reducing the block size from 1024 to 128 lowers encoder FLOPs substantially, to roughly 30\% of the value at 1024, whereas the measured encoder latency decreases only to about 77\%. This gap indicates that current GPU kernels do not yet fully translate the reduced attention cost of many small blocks into proportional wall-clock speedups, suggesting additional systems-level headroom for future optimization.


\vspace{1mm}
\noindent\textbf{Decoder sectorization and encoder supervision.}
Table~\ref{tab:ablation} ablates the decoder sectorization strategy and the auxiliary encoder supervision used in HEPTv2. The first group of results shows that sectorized decoding is essential: replacing the full-event decoder with a sectorized decoder improves \ac{DM} from 96.9\% to 98.2\% with two sectors and to 98.6\% with three sectors, while keeping fake rate essentially unchanged. Increasing the number of sectors beyond three provides no further benefit and may slightly worsen fake rate, indicating that a small number of broad \(\phi\)-sectors are sufficient to capture the approximate locality structure of the assignment problem. 
The remaining ablations show that auxiliary encoder supervision is also important. 
Removing encoder supervision from the default three-sector decoder reduces \ac{DM} from 98.6\% to 96.3\%, confirming that direct hit-to-track assignment benefits substantially from encoder features that are jointly trained both to distinguish target hits from background and to produce more discriminative hit representations. 
We also evaluated a variant with overlapping sectors by augmenting the default three non-overlapping \(\phi\)-sectors with a second set of three shifted sector windows, yielding six sector windows in total. However, this configuration produced limited measurable gain. This is likely because only a very small fraction of tracks are affected by sector boundaries when using three sectors: in our experiments, even if all such boundary failures were resolved perfectly, the attainable improvement in \ac{DM} would be only about \(0.1\%\). HEPTv2 therefore adopts non-overlapping sectors by default, avoiding additional decoding cost without sacrificing tracking quality.

\section{Discussion}
\label{sec:discussion}

\vspace{1mm}
\noindent\textbf{Implications for graph-based and transformer-based tracking pipelines.}
The comparison with both graph-based and transformer-based baselines suggests that the central ingredient in learning-based tracking is not the graph representation itself, but the ability to preserve and exploit locality under practical computational constraints. Strong GNN pipelines inject locality explicitly through graph construction and message passing, whereas naive transformer formulations risk allocating much of their computation to irrelevant hit pairs. HEPTv2 shows that LSH-based serialization and sectorized decoding can recover this essential inductive bias within a more hardware-efficient transformer pipeline. At the same time, the gains over HEPT+DBSCAN and two-stage MF indicate that neither locality-aware encoding alone nor generic direct decoding alone is sufficient. Instead, the strongest performance arises when locality-aware hit processing is coupled with end-to-end track construction under a unified training objective.

\vspace{1mm}
\noindent\textbf{Scope and future extensions.}
This study evaluates HEPTv2 on the TrackML pixel-detector benchmark~\cite{amrouche2019tracking,kiehn2019trackml}, which provides a controlled setting for studying scalable hit-to-track assignment in high-occupancy charged-particle tracking. This setting retains the central combinatorial structure of the tracking problem and enables direct comparison with prior learning-based approaches. Extending HEPTv2 to full-detector inputs, including both pixel and strip measurements, experiment-specific geometries, downstream ambiguity resolution, and track fitting, is an important direction for future work. 
Future work should also examine robustness under broader detector effects and higher pile-up, and study detailed failure modes such as merged, split, and duplicate tracks. On the systems side, further optimization through quantization~\cite{sun2024flatquant,lin2025qserve}, fused kernels, compiler-level optimization~\cite{zhai2023bytetransformer}, and other hardware-aware improvements could further improve latency and memory efficiency.  

More broadly, HEPTv2 points to a design principle for scientific machine learning on large irregular measurement sets: task-specific locality can be combined with end-to-end transformer-based reconstruction to identify latent physical objects from unordered point clouds. Many scientific reconstruction problems involve sparse, noisy, non-uniform measurements in which object identity must be inferred without a regular grid or a fixed graph structure. In such settings, explicit graph construction can encode useful locality but may become a computational bottleneck, whereas generic global attention is often too costly and insufficiently structured. The combination of locality-preserving serialization, region-wise decoding, and end-to-end assignment learning offers a route for transforming clustering-like reconstruction problems into hardware-efficient transformer pipelines. This principle may be relevant not only to charged-particle tracking, but also to other HEP and astroparticle-physics tasks such as machine-learned particle-flow reconstruction~\cite{pata2021mlpf}, detector readout and object reconstruction in high-granularity or liquid-argon detectors~\cite{qasim2019learning,domine2020scalable}, and sparse event reconstruction in neutrino telescopes such as IceCube~\cite{abbasi2022graph,soegaard2022graphnet}.

\section{Methods}
\label{sec:methods}

\subsection{TrackML Dataset and Task Formulation}\label{sec:methods:trackml}

Our evaluation is based on the TrackML dataset~\cite{amrouche2019tracking,kiehn2019trackml}, a standard high-energy physics benchmark for charged-particle track reconstruction. TrackML provides a controlled setting for studying high-occupancy collider tracking, using a generic silicon tracking detector model motivated by the challenges expected at the \ac{HL-LHC}. The simulated events consist of \(t\bar{t}\) hard scatterings overlaid with soft QCD interactions. Each event contains simulated charged particles that, as they traverse successive detector layers, leave sparse trajectories of spatial measurements (detector hits) across the tracking detector. The reconstruction task is to associate the hits produced by the same particle into a trajectory, as illustrated in Figure~\ref{fig:heptv2}. Following recent TrackML-based studies~\cite{lieret2023high,miao2024locality,van2025transformers}, we consider the pixel-detector setting of the benchmark. This setting yields on the order of \(6\times10^4\) hits per event, making efficient reconstruction particularly challenging. Reconstructable target particles are defined as charged particles with \(\pt>0.9~\mathrm{GeV}\) and \(|\eta|<4\) that leave at least three hits in the pixel detector, following the standard selection~\cite{lieret2023high,miao2024locality,acorngnn}. Hits produced by particles outside this target definition, together with detector noise hits that are not associated with any target particle, are treated as background.

Formally, for an event with \(N\) detector hits, let $K$ denote the number of reconstructable target particles. Its ground-truth hit-to-track assignment can then be represented by a binary matrix \(A^\star\in\{0,1\}^{K\times N}\), whose entries are defined as
\[
A^\star_{k,i}=
\begin{cases}
1, & \text{if hit } i \text{ was produced by target particle } k,\\
0, & \text{otherwise.}
\end{cases}
\]
Charged-particle tracking is therefore formulated as reconstructing \(A^\star\) from the unordered set of detector hits.


\subsection{HEPTv2 Architecture}
\label{sec:methods:heptv2}

HEPTv2 combines a locality-aware point encoder with a sectorized decoder to reconstruct charged-particle tracks directly from unordered detector hits in a single end-to-end pipeline, as illustrated in Figure~\ref{fig:heptv2}. We describe these components and the corresponding training objectives below.

\subsubsection{Locality-Aware Point Encoder}

The encoder of HEPTv2 is designed to produce contextualized hit representations efficiently at full-event scale, so that downstream track construction can operate on features that already encode the local geometric structure of charged-particle trajectories. 
To do so, 
HEPTv2 builds on a point-serialization idea of HEPT that is well aligned with tracking geometry. 
In its previous version, HEPT, LSH is applied to latent query/key representations and adopts a custom \(L_2\)-distance-based attention kernel.  
In HEPTv2, by contrast, serialization is performed directly in detector coordinate space, specifically in \((\eta,\phi)\), as experiments show that Euclidean proximity in this space provides a faithful proxy for tracking locality\footnote{In principle, alternative hashing schemes defined in raw 3D space, such as angular-based LSH, may also be viable, but we leave that direction to future work.}. This decouples the serialization mechanism from the attention kernel, enabling the use of scaled dot-product attention with mature hardware and kernel support.

Concretely, for each hit \(i\), we extend the OR--AND LSH serialization~\cite{miao2024locality} to the detector-space coordinate \((\eta_i,\phi_i)\), obtaining a scalar ordering value \(o_i=\mathrm{LSH}_{\mathrm{HEPT}}(\eta_i,\phi_i)\). 
Conceptually, \(o_i\) is constructed from \(m_{\mathrm{OR}}\) independent hash tables, each using \(m_{\mathrm{AND}}\) E\(^2\)LSH projections~\cite{datar2004locality}: the AND rule in $\mathrm{LSH}_{\mathrm{HEPT}}$ requires two hits to be nearby under all \(m_{\mathrm{AND}}\) projections within a table to reduce accidental collisions between unrelated hits, while the OR rule collects the union of the results of \(m_{\mathrm{OR}}\) tables so that proximity in any one table can keep the two hits close in the final ordering, thereby improving recall of true local neighbors.
Therefore, sorting the hits by \(o_i\) yields a one-dimensional sequence that probabilistically preserves locality in the original \((\eta,\phi)\) space. The sequence is then partitioned into contiguous blocks of fixed size, and self-attention is restricted to hits within the same block, resulting in a block-diagonal attention pattern with cost linear in \(N\). To reduce boundary artifacts from any single serialization, independent OR--AND LSH projections are used across attention heads within each encoder layer, so that the same event is essentially viewed through multiple randomized serialized orders.

Let \(X\in\mathbb{R}^{N\times d}\) denote the input hit features after the input projection layer. HEPTv2 adopts a pre-normalized transformer block with per-head query/key normalization for each of its encoder layer: Specifically, let \(\tilde{X}=\mathrm{RMSNorm}(X)\), and for each attention head $h$, let \(Q_h=\mathrm{RMSNorm}(\tilde{X}W_h^Q)\), \(K_h=\mathrm{RMSNorm}(\tilde{X}W_h^K)\), and \(V_h=\tilde{X}W_h^V\). Scaled dot-product attention \(\mathrm{Attn}_h(\tilde{X})=\mathrm{softmax}\!\left(Q_hK_h^\top/\sqrt{d_h}\right)V_h\) is then computed within each serialized block, where $d_h$ is the dimension for each head.

After the last encoder layer, the encoder outputs hit representations \(H\in\mathbb{R}^{N\times d}\).
Unless otherwise stated, the encoder uses 4 layers, 8 attention heads with head dimension 128, a block size of 1024, and an \(\mathrm{LSH}_{\mathrm{HEPT}}\) configuration with $m_{\mathrm{OR}}=3$ and $m_{\mathrm{AND}} = 2$.

\subsubsection{Sectorized Track Decoder}

The decoder of HEPTv2 is designed to directly predicts the hit-to-track assignment from the hit representations produced by the encoder.
Because the number of reconstructable tracks varies across events, HEPTv2 maintains a fixed bank of \(M\) learnable latent vectors, each representing one track slot, denoted by
$
T^{\mathrm{init}} \in \mathbb{R}^{M\times d},
$
where each row of \(T^{\mathrm{init}}\) is a \(d\)-dimensional track slot representation and \(M\) is chosen to exceed the maximum expected number of reconstructable tracks in the events of interest.
In any given event, only a subset of these slots becomes active, while the remainder are learned to be inactive, allowing the number of active slots to adapt to the number of tracks present.
Under this formulation, track reconstruction amounts to predicting which track slots should become active and which hits should be assigned to each active slot. 
The decoder therefore predicts both a binary classification vector \(\hat{y}\in\mathbb{R}^{M}\), whose \(m\)-th entry indicates whether track slot \(m\) should be active or inactive, and an assignment matrix \(\hat{A}\in\mathbb{R}^{M\times N}\), whose rows correspond to track slots, columns correspond to detector hits, and entries give learned assignment scores for associating individual hits with individual slots.

However, a direct full-event prediction of the \(M\times N\) hit-to-track assignment is unnecessarily challenging because of the high ambiguity present in high-pileup events.
HEPTv2 therefore partitions the encoded event into \(k\) broad \(\phi\)-sectors  and solves \(k\) smaller assignment subproblems, each of approximate size \(M\times (N/k)\); unless otherwise stated, we use \(k=3\) in HEPTv2, corresponding to an azimuthal sector width of \(\Delta\phi=2\pi/3\).
This exploits the approximate block-sparse structure of the ground-truth global hit-to-track assignment matrix \(A^\star\), as illustrated in Figure~\ref{fig:heptv2} (bottom right): for a given track, valid hit assignments are typically concentrated within a certain \(\phi\) region, so one large full-event assignment can be replaced by several smaller and less ambiguous sector-level problems.
Thus, for each sector \(s\), let \(H_s\in\mathbb{R}^{N_s\times d}\) denote the encoded hit representations of the \(N_s\) hits in that sector. The decoder now predicts two quantities for each sector: a slot classification output \(\hat{y}_s\in\mathbb{R}^{M}\), indicating which of the \(M\) track slots should be active for that sector, and a hit-assignment matrix \(\hat A_s\in\mathbb{R}^{M\times N_s}\), specifying how the \(N_s\) hits in that sector should be assigned to those track slots.

To produce these sector-level outputs \(\hat y_s\) and \(\hat A_s\), HEPTv2 iteratively refines the track-slot representations against the encoded hit representations \(H_s\) in sector \(s\). Starting from \(T_s^{(0)}=T^{\mathrm{init}}\), the sector-specific slot states \(T_s^{(\ell)}\in\mathbb{R}^{M\times d}\) are updated across \(L\) decoder layers. In each layer, cross-attention allows every track slot to gather evidence from the sector hits, so that each slot can look over available hits and assign higher scores for those hits that best support its  trajectory hypothesis. Self-attention then lets the slots exchange information with one another so that competing or duplicate trajectory hypotheses can be suppressed. Formally, the layer update is
\[
T_{s,\mathrm{cross}}^{(\ell)}
=
\mathrm{CrossAttn}\!\left(T_s^{(\ell)},\, H_s\right),
\quad
T_{s,\mathrm{self}}^{(\ell)}
=
\mathrm{SelfAttn}\!\left(T_{s,\mathrm{cross}}^{(\ell)}\right),
\quad
T_s^{(\ell+1)}
=
\mathrm{FFN}\!\left(T_{s,\mathrm{self}}^{(\ell)}\right).
\]
After the final decoder layer \(L\), a classification head predicts whether a slot is active, \(\hat{y}_s=\mathrm{MLP}_{\mathrm{cls}}(T_s^{(L)})\in\mathbb{R}^{M}\), and an assignment head produces the final slot embeddings, \(E_s=\mathrm{MLP}_{\mathrm{assign}}(T_s^{(L)})\in\mathbb{R}^{M\times d}\). The corresponding sector-level hit-assignment logits are then obtained as
\[
\hat{A}_s = E_s H_s^\top \in \mathbb{R}^{M\times N_s},
\]
whose \((m,n)\)-th entry gives the assignment score between track slot \(m\) and hit \(n\) in sector \(s\), and $\hat{y}_s$ tells which rows of $\hat{A}_s$ are active slots.

Finally, when a track crosses a sector boundary, its hits may be split across multiple sectors, so that a sector-level decoder sees only a partial trajectory and may produce a fragmented prediction. To mitigate this effect, we also evaluated an overlapping-sector variant in which the default \(k\) non-overlapping \(\phi\)-sectors are supplemented by a second set of \(k\) shifted sector windows, offset by half a sector width. For the default choice \(k=3\), this yields six sector windows in total. The purpose of this shifted partition is to ensure that a track lying near a boundary in the original partition is likely to lie well inside one of the shifted sectors. 
As shown in Table~\ref{tab:ablation}, however, the empirical gain from this overlap scheme is minimal. We further investigated this effect and found that, compared against the default choice \(k=3\), even perfect recovery of all boundary-crossing tracks would improve \ac{DM} by only about 0.1\%. We therefore adopt the simpler non-overlapping design and directly merge the sector-level assignment predictions \(\hat{A}_s\) to obtain the final event-level reconstruction \(\hat{A}\). Unless otherwise stated, the decoder uses \(L=2\) decoder layers and \(M=3000\) learnable track slots.

\subsubsection{Training Objectives}
The locality-aware encoder and the sectorized track decoder in HEPTv2 are trained end to end with a unified objective that jointly optimizes for the same final tracking task. 
On the encoder side, auxiliary supervision encourages the model to distinguish target hits from background and to learn discriminative hit representations. On the decoder side, supervision is applied directly to hit assignment, so that the learned track slots are trained to recover the sector-level hit-to-track assignment.

\vspace{1mm}
\noindent\textbf{Encoder-side supervision.}
Let \(H=[h_1,\ldots,h_N]^\top\in\mathbb{R}^{N\times d}\) denote the final hit representations produced by the encoder. 
From each \(h_i\), we derive two auxiliary outputs, \(z_i=h_i/\|h_i\|_2\) and \(b_i=\mathrm{Linear}(h_i)\), where \(z_i\in\mathbb{R}^{d}\) is the \(\ell_2\)-normalized embedding used for contrastive supervision and \(b_i\in\mathbb{R}\) is the hit-level logit for target-versus-background classification. The contrastive supervision encourages hits from the same target trajectory to remain close in the embedding space while pushing hits from different trajectories apart, whereas the classification enables the encoder to better distinguish target hits from background.

Specifically, for a target hit \(i\), let the positive set \(\mathcal{P}(i)\) include the collection of other hits produced by the same target particle, and let the negative set \(\mathcal{N}(i)\) include the collection of \(K_{\mathrm{neg}}\) hardest negatives, corresponding to the hits from other particles or background with the largest cosine similarity to \(z_i\). We apply an InfoNCE loss
\[
\mathcal{L}_{\mathrm{InfoNCE}}
=
-
\log
\frac{
\sum_{j\in\mathcal{P}(i)} \exp(z_i^\top z_j/\tau)
}{
\sum_{j\in\mathcal{P}(i)} \exp(z_i^\top z_j/\tau)
+
\sum_{a\in\mathcal{N}(i)} \exp(z_i^\top z_a/\tau)
},
\]
where \(\tau\) is the temperature. Then, we supervise the background logits with a binary cross-entropy loss,
$
\mathcal{L}_{\mathrm{bg}}
=
\frac{1}{N}\sum_{i=1}^{N} \mathrm{BCELogit}(b_i,b_i^{\star}),
$
where \(b_i^\star=1\) if hit \(i\) is a background hit, i.e., it does not belong to any target particle, and \(b_i^\star=0\) otherwise. Unless otherwise stated, we use \(K_{\mathrm{neg}}=256\) and \(\tau=0.07\).

\vspace{1mm}
\noindent\textbf{Decoder-side supervision.}
Decoder-side supervision has two goals: to determine which track slots should become active and to train each active slot to recover the correct hit assignment. 
For sector \(s\), let \(A_s^\star\in\{0,1\}^{K_s\times N_s}\) denote the ground-truth hit-to-track assignment matrix over the \(N_s\) hits and \(K_s\) target tracks in that sector.
The decoder predicts two corresponding outputs: sector-level hit-assignment logits \(\hat A_s\in\mathbb{R}^{M\times N_s}\), and binary classification logits \(\hat y_s\in\mathbb{R}^{M}\), whose elements indicate which of the \(M\) track slots are active in that sector.

Because the ordering of the learned track slots is arbitrary and \(M\) may exceed \(K_s\), the correspondence between the rows of \(\hat A_s\in\mathbb{R}^{M\times N_s}\)  and those of \(A_s^\star\in\{0,1\}^{K_s\times N_s}\) is not known a priori. We therefore use the Hungarian matching algorithm~\cite{kuhn1955hungarian} to align the \(K_s\) ground-truth tracks with the best corresponding \(K_s\) rows of \(\hat A_s\), yielding \(\bar A_s\in\mathbb{R}^{K_s\times N_s}\) from \(\hat A_s\).
Those matched  \(K_s\) rows of \(\hat A_s\) also determine which track slots are active, thereby providing binary cross entropy supervision labels \(y_s^\star\in\{0,1\}^{M}\) for \(\hat y_s\); and the corresponding loss function is denoted as $\mathcal{L}_{\mathrm{cls}}^{(s)}$.

The matched assignment matrix \(\bar A_s\in\mathbb{R}^{K_s\times N_s}\) is then supervised against the ground-truth assignment matrix \(A_s^\star\in\{0,1\}^{K_s\times N_s}\) using two complementary losses. First, a standard sigmoid focal loss~\cite{lin2017focal}, denoted as $\mathcal{L}_{\mathrm{assign}}^{(s)}$, is applied elementwise to the logits in \(\bar A_s\) and the corresponding binary targets in \(A_s^\star\), encouraging the model to focus on difficult hit-assignment decisions under strong class imbalance. Second, a Dice overlap loss~\cite{milletari2016v} is applied row-wise, so that each predicted track-level hit assignment is encouraged to overlap well with the corresponding ground-truth track. Specifically, the Dice overlap loss is
\[
\mathcal{L}_{\mathrm{dice}}^{(s)}
=
\frac{1}{K_s}\sum_{j=1}^{K_s}
\left(
1-\frac{2\,\hat{a}_j^\top a_j+1}{\|\hat{a}_j\|_1+\|a_j\|_1+1}
\right),
\]
where \(\hat{a}_j\in[0,1]^{N_s}\) and \(a_j\in\{0,1\}^{N_s}\) denote the \(j\)-th rows of \(\operatorname{Sigmoid}(\bar A_s)\) and \(A_s^\star\), respectively.  This loss directly measures the overlap between the predicted and ground-truth hit sets for each track, thereby complementing the element-wise focal loss with a set-level supervision signal.

\vspace{1mm}
\noindent\textbf{Optimization details.}
For each training event, the overall objective combines the encoder-side losses with the decoder-side losses averaged over the \(k\) sectors:
\[
\mathcal{L}
=
\lambda_{\mathrm{InfoNCE}}\mathcal{L}_{\mathrm{InfoNCE}}
+
\lambda_{\mathrm{bg}}\mathcal{L}_{\mathrm{bg}}
+
\frac{1}{k}
\sum_{s=1}^{k}
\left(
\lambda_{\mathrm{cls}}\mathcal{L}_{\mathrm{cls}}^{(s)}
+
\lambda_{\mathrm{assign}}\mathcal{L}_{\mathrm{assign}}^{(s)}
+
\lambda_{\mathrm{dice}}\mathcal{L}_{\mathrm{dice}}^{(s)}
\right).
\]
Unless otherwise stated, we use \(\lambda_{\mathrm{cls}}=0.1\), \(\lambda_{\mathrm{assign}}=200\), \(\lambda_{\mathrm{dice}}=2\), \(\lambda_{\mathrm{bg}}=1.8\), and \(\lambda_{\mathrm{InfoNCE}}=12\). We train the model for 250 epochs with batch size 1 using the Muon optimizer~\cite{liu2025muon}, with learning rate set to \(2.5\times10^{-4}\). The learning rate is decayed with a step schedule of size 25 and decay factor 0.5.

\subsection{Baseline Methods and Evaluation Protocol}
\label{sec:methods:baselines}

\subsubsection{Baseline Methods}

We compare HEPTv2 with four representative learning-based baselines: the GNN-based pipelines ACORN-GNN~\cite{acorngnn} and OC-GNN~\cite{lieret2023high}, and the transformer-based approaches HEPT+DBSCAN~\cite{miao2024locality} and Two-stage MF~\cite{van2025transformers}.

\vspace{1mm}
\noindent\textbf{ACORN-GNN}~\cite{ju2021performance,acorngnn} is an edge-classification pipeline for particle tracking. It first uses a metric-learning network to embed detector hits into a latent space, constructs candidate edges by connecting nearby hits in that space, and applies a filter network to prune the resulting graph. A GNN then classifies the remaining hit-to-hit edges, and graph-level post-processing converts the accepted edges into track candidates using the FastWalkthrough algorithm~\cite{atlas2024computational}. This baseline represents a strong graph-based tracking paradigm, but requires multiple potentially expensive stages---metric learning, graph construction, edge filtering, GNN edge classification, and graph segmentation---before final tracks are produced. Each of these steps is optimized separately to reduce graph size and achieve the target physics performance. 

\vspace{1mm}
\noindent\textbf{OC-GNN}~\cite{lieret2023high} follows the object-condensation paradigm for high-pileup particle tracking. It first constructs a sparse graph in the latent space learned by a metric-learning model. A GNN is then applied to update hit representations and predict object-condensation quantities, including per-hit condensation coordinates and condensation scores. These outputs encourage hits from the same charged particle to cluster around common condensation centers in a learned latent space, after which final track candidates are obtained through a separate DBSCAN clustering step. Compared with ACORN-GNN, OC-GNN removes the explicit edge-pruning and edge-classification stages, but still depends on expensive graph construction and post-hoc clustering.

\vspace{1mm}
\noindent\textbf{HEPT+DBSCAN}~\cite{miao2024locality} combines the original HEPT encoder with density-based clustering. HEPT uses LSH to serialize nearby points into regular attention blocks, enabling transformer layers to preserve local geometric structure while using local-window attention to avoid the quadratic cost of global attention. In its original tracking formulation, HEPT produces per-hit embeddings that are trained so that hits from the same particle are close in the learned feature space. To obtain full track candidates, HEPT+DBSCAN applies DBSCAN to these learned hit embeddings as a post-hoc clustering step, analogous to the clustering stage used in OC-GNN.

\vspace{1mm}
\noindent\textbf{Two-stage MF}~\cite{van2025transformers} is a two-stage transformer-based tracking baseline that adapts MaskFormer~\cite{cheng2022masked} to charged-particle reconstruction. 
In the first stage, a transformer is trained to filter background hits and thereby reduce the input size for the second stage, where a MaskFormer-style decoder predicts a per-track hit-assignment matrix over the remaining hits, together with corresponding track-confidence outputs. 
The hit-filtering and track-construction modules are therefore optimized separately under different objectives. 
To control computational cost, the model orders hits by azimuthal angle \(\phi\) and applies sliding-window attention over the resulting sequence. 
This provides a simple way to impose locality, but the locality structure is tied to a fixed one-dimensional ordering of the detector hit cloud. 
As event size grows, such a coarse ordering can place unrelated hits within the same local window while separating hits that belong to the same trajectory, reducing the efficiency with which attention is allocated and making accurate track reconstruction more difficult.

\subsubsection{Evaluation protocol}
\label{sec:methods:evaluation}

\noindent\textbf{Tracking metrics.}
For tracking quality, we report Double Majority (DM) efficiency and fake track rate, following the evaluation protocol used in recent TrackML-based studies~\cite{lieret2023high,miao2024locality,van2025transformers}. Let \(\mathcal{T}\) denote the set of reconstructable target particles in an event and \(\mathcal{R}\) denote the set of reconstructed track candidates with at least three hits. A reconstructed track \(r\in\mathcal{R}\) is considered to match a target particle \(t\in\mathcal{T}\) if it satisfies the double-majority criterion,
\[
\frac{|\mathrm{Hits}(r)\cap \mathrm{Hits}(t)|}{|\mathrm{Hits}(r)|} > 0.5
\qquad \text{and} \qquad
\frac{|\mathrm{Hits}(r)\cap \mathrm{Hits}(t)|}{|\mathrm{Hits}(t)|} > 0.5,
\]
where \(\mathrm{Hits}(r)\) and \(\mathrm{Hits}(t)\) denote the sets of detector hits associated with \(r\) and \(t\), respectively. The DM efficiency and fake track rate are then defined as
\[
\epsilon_{\mathrm{DM}}
=
\frac{
\left|\{t\in\mathcal{T}: \exists r\in\mathcal{R}\ \text{such that } r \text{ matches } t\}\right|
}{|\mathcal{T}|},
\qquad
f_{\mathrm{DM}}
=
\frac{
\left|\{r\in\mathcal{R}: \nexists t\in\mathcal{T}\ \text{such that } r \text{ matches } t\}\right|
}{|\mathcal{R}|}.
\]


\noindent\textbf{Computational efficiency.} We evaluate computational efficiency using end-to-end inference latency and peak GPU memory usage on a single NVIDIA A100 GPU for HEPTv2 and the GNN baselines, ACORN-GNN and OC-GNN, using their official implementations.
Latency is measured in milliseconds per event from preprocessed event-level hit tensors to final reconstructed track assignments. We exclude disk I/O and data loading, but include all method-specific steps required to produce final tracks, such as graph construction, clustering, hit filtering, mask decoding, neural-network inference, and post-processing when applicable. This protocol is intended to reflect the cost of an online-inference setting after event data have been loaded into memory. For HEPTv2 and baselines profiled in our implementation, we measure latency with \texttt{torch.utils.benchmark}. Each method is run in evaluation mode with a batch size of one event. We perform 20 warm-up iterations before timing, apply CUDA synchronization immediately before and after the timed region, and report the average latency over the evaluation set. Peak GPU memory is measured over the same inference procedure by resetting the PyTorch peak-memory counters before inference and reading \texttt{torch.cuda.max\_memory\_allocated()} after synchronization. We report the maximum peak allocation observed across evaluated events.

\section*{Data Availability}
Datasets used in this study are freely available on Kaggle at \hyperlink{https://www.kaggle.com/competitions/trackml-particle-identification/data}{https://www.kaggle.com/competitions/trackml-particle-identification/data} \cite{kiehn2019trackml}.

\section*{Code Availability}
The source code for this study is publicly available on GitHub at \hyperlink{https://github.com/Graph-COM/HEPTv2}{https://github.com/Graph-COM/HEPTv2}. The ACORN-GNN configuration can be found at \hyperlink{https://github.com/jackrodgers3/HEPT\_GNN\_PIPELINE}{https://github.com/jackrodgers3/HEPT\_GNN\_PIPELINE}.

\section*{Acknowledgments}
S.M., S.G., and P.L. were supported by the NSF (National Science Foundation) under grant numbers PHY-2117997 and IIS-2435957. S.M., S.G., and P.L. were also supported by the Georgia Institute of Technology IDEaS Cyberinfrastructure Awards. Y.C., S.H., J.P.R., M.L., and J.D. are also supported by the NSF grant No. PHY-2117997. This research used resources of the National Energy Research Scientific Computing Center, a U.S. Department of Energy Office of Science User Facility supported by the Office of Science of the U.S. Department of Energy under Contract No. DE-AC02-05CH11231 using NERSC award NERSC DDR-ERCAP0034643. 

\section*{Author Contributions Statement}
S.M., Y.C., and P.L. conceived the project. S.M., S.G., J.P.R., and Y.C. developed the code. S.M., S.G., and J.P.R. conducted the experiments. S.M., S.G., Y.C., and P.L. wrote the manuscript. M.L., J.D., S.H., Y.C., and P.L. provided supervision.

\section*{Competing Interests Statement}
We declare that none of the authors have competing financial or non-financial interests as defined by Nature Portfolio.

\bibliography{references,new_ref}
\end{document}